\documentclass[runningheads]{llncs}
\usepackage{graphicx}
\usepackage{multirow}
\begin{document}
%
\title{Political Framing: US COVID19 Blame Game}
%
%
\author{Chereen Shurafa\inst{1} \and
Kareem Darwish\inst{2} \and
Wajdi Zaghouani\inst{1}}
\authorrunning{C. Shurafa et al.}
%
\institute{$^1$College of Humanities and Social Sciences, HBKU, Qatar 
\email{chereenalshurafa@gmail.com, wzaghouani@hbku.edu.qa}\\
$^2$Qatar Computing Research Institute, HBKU, Qatar\\
\email{kdarwish@hbku.edu.qa}
}
\maketitle              
\begin{abstract}
Through the use of Twitter, framing has become a prominent presidential campaign tool for politically active users. Framing is used to influence thoughts by evoking a particular perspective on an event. 
In this paper, we show that the COVID19 pandemic rather than being viewed as a public health issue, political rhetoric surrounding it is mostly shaped through a blame frame (blame Trump, China, or conspiracies) and a support frame (support candidates) backing the agenda of Republican and Democratic users in the lead up to the 2020 presidential campaign. We elucidate the divergences between supporters of both parties on Twitter via the use of frames. 
Additionally, we show how framing is used to positively or negatively reinforce users’ thoughts. We look at how Twitter can efficiently be used to identify frames for topics through a reproducible pipeline. 

\keywords{Framing Theory  \and COVID19 \and Social Media Analysis}
\\

\textit{This paper was accepted at SocInfo2020.  Please cite the SocInfo version.} 
\end{abstract}
\section{Introduction}
The US has been the hardest hit country by the COVID19 pandemic with 100,000 deaths as of May 26, 2020. Though the pandemic is a public health problem, the response to the pandemic among politically active US Twitter users took a political turn due to diverging views between Republicans and Democrats.  Both sides turned to actively placing blame on different entities for the spread of COVID19 through the use of carefully crafted phrases. For the Democrats, the effect is manifested through blaming Trump for his inability to act quickly and protect the nation from the rapidly spreading virus. Conversely, the Republicans rallied together to protect Trump and consistently shifted the blame to deep state conspiracies and to China for being the source of the pandemic.  
In this paper, we examine the divergence between supporters of the Republican (GOP) and Democratic (DNC) parties\footnote{GOP: \textit{Grand Old Party}; DNC: \textit{Democratic National Committee}. } in the context of framing theory, where frames shape how individuals construct their attitudes towards an issue~\cite{chong2007framing}.  Using an expectancy value model, an individual's overall attitude towards an issue is determined using a weighted average of different factors of an issue, where the weights correspond to the relative importance of these factors to the individual~\cite{nelson1997toward}. For the COVID19 pandemic, different factors may include perceived political responsibility, the state of the economy, the accessibility of health care, and effect of the pandemic on elections.  In effect, a person's attitude is framed by the perceived importance of such factors.  Framing works at three levels, namely the existing of factors in memory, meaning if the person is aware of the factor, whether this factor is mentally accessible when forming an attitude, and the perceived relevance of the factor to the situation~\cite{chong2007framing}.  We look at how framing is affecting the Twitter political discourse associated with the COVID19 pandemic.  We base our analysis on the timeline tweets of 30 thousand Twitter users who discussed the pandemic in conjunction with politically suggestive words.  We utilize unsupervised stance detection to automatically and accurately distinguish between supporters of the Democratic and Republican parties. We establish the salient background topics these users have discussed in the period between Jan. 1, 2020 and April 12, 2020, which includes the period leading to the pandemic and the period after the World Health Organization (WHO) declared COVID19 as a global pandemic.  We contrast politically active users against a set of 20 thousand random US users.  The contributions of this paper are as follows:
\begin{itemize}
    \item We illustrate how framing can be used to shape the public discourse.  Specifically, we show that supporters of the two main political parties in the US are framing the COVID19 pandemic as a political issue.
    \item We utilize multiple analysis tools ranging from stance detection to detect political leanings of users to conducting an analysis based hashtags, URLs, retweeted accounts, and rhetorical devices.
    \item We show that our analysis of different discourse elements can help identify underlying frames.
\end{itemize}


%
%
%

\section{Related Work}
Related research on political discourse framing is well established in the social science community as described in \cite{chong2007framing,entman1993framing} 
and also by automatically analyzing political discourse and news articles using natural language processing techniques \cite{baumer2015testing,card2015media}. 
Other computational works focused on analyzing social media political discourse using opinion mining  \cite{abu2013identifying}  
and stance detection \cite{ebrahimi2016joint,hasan2014you,johnson2016identifying,sridhar2015joint,walker2012stance}. 
Stance detection is the task of identifying the political leaning or the position of a user towards a specific topic or entity. In the context of Twitter, User stances can be gleaned using a myriad of signals including content features (e.g. words and hashtags), profile information (e.g. profile description), interactions (e.g. retweets, likes, and mentions) \cite{aldayel2019your}. There are multiple methods for performing stance detection including supervised, semi-supervised, and unsupervised methods~\cite{borge2015content,darwish2018scotus,darwish2019unsupervisedStance}. In this work, we employ unsupervised stance detection due in most part to its ease and high accuracy~\cite{darwish2019unsupervisedStance}.

Framing in public discourse such as in political speeches and
news articles have been explored by ~\cite{baumer2015testing,card2015media,fulgoni-etal-2016-empirical}. While ~\cite{recasens-etal-2013-linguistic,choi-etal-2012-hedge,greene-resnik-2009-words} discussed framing in the context of language bias and subjectivity ~\cite{wiebe2004}.
On the other hand, ~\cite{tan-etal-2014-effect} analyzed the effect of lexical choices in Twitter and how they can affect the propagation of the tweets. ~\cite{pla-hurtado-2014-political,bakliwal-etal-2013-sentiment} studied sentiment analysis of the political discourse and ~\cite{sim-etal-2013-measuring,djemili2014does} created a framework to measure and predict ideologies.  \cite{Gerrish,bermingham-smeaton-2011-using} analyzed the voting patterns and polls based on political sentiment analysis. Furthermore, several social sciences researchers have discussed framing based on Twitter data and how it has been used to impact the public opinion \cite{jang2015,Burch2015,Harlow2011}. For instance, ~\cite{Groshek} discussed the framing of Public sentiment in social media content during the 2012 U.S. presidential campaign, and ~\cite{Pond2019} presented his research on social media and framing discourse in the digital public sphere based on riots and digital activism. Finally, ~\cite{johnson-etal-2017-ideological} presented a model based on the tweet's linguistic features and several ideological phrase indicators to predict the general frame of political tweets.

\section{Data Collection}
\textbf{\textit{Obtaining Tweets}}
Our target was to assemble a dataset composed of Twitter users who discuss COVID19. Specifically, we were interested in users who politically lean towards the Republican (GOP) or the Democratic (DNC) parties and, for reference, a random sample of US users.  We collected tweets from the period of March 5 -- 31, 2020 using the 
Twitter streaming API.  We filtered the tweets by language, to retain English tweets, and by the following hashtags and keywords: \#covid19, \#CoronavirusOutbreak, \#Coronavirus, \#Corona, \#CoronaAlert, \#CoronaOutbreak, Corona, and COVID19.  To filter by language, we used the language tag provided by Twitter for each tweet.  In all, we obtained a set of 31.64 million tweets, which we henceforth refer to as the \textit{base} dataset.

To identify users with interest in US politics, we further filtered tweets containing any of the following strings (or potentially sub-strings): Republican, Democrat, Trump, Biden, GOP, DNC, Sanders, and Bernie, which resulted in 2.48 million tweets.  We sorted in descending order users by the number of tweets that mention COVID19 related words and politically indicative words and retained the top 30k users. These users had tweets that ranged in number between 1 and 1,302 tweets (average: 2.78 tweets and standard deviation: 6.72).  We proceeded to crawl the timeline tweets for these 30k users.  Twitter APIs allow the scrapping of the last 3,250 timeline tweets for a user.  From timeline tweets, we retained tweets with dates ranging between January 1 and April 12, 2020, which were 92.02 million tweets. Of those, 18.55 million tweets contained the substrings \textit{corona} or \textit{covid}, and the users had tweets ranging in number between 1 and 3,250 tweets (average: 677.52 tweets and standard deviation: 292.88).  We shall refer to this dataset as the \textit{politicised} dataset.

To sample US users in general, we filtered all users with tweets in the \textit{base} dataset by their locations to obtain US users.  We deemed a user to be from the US if their specified location (from their Twitter profile) contained the tokens ``United States", ``America'', ``USA'', or the names of any state or its abbreviation (e.g. ``Maryland'' or ``MD'').  For USA and state abbreviations, they had to be capitalized.  From the 117,408 matching users, we randomly sampled 20k users and obtained their timeline tweets.  In the period of interest (Jan. 1 -- April 12, 2020), these users had 14.12 million tweets, of which 773k tweets contained the substrings \textit{corona} or \textit{covid}.  We refer to this as the \textit{sampled} dataset.

One obvious difference between the \textit{politicised} and the \textit{sampled} datasets is that the users in the politicised dataset produced more than 4.3 times more tweets in general and 16 times more COVID19 related tweets.  The difference may be an artifact of identifying the 30k most politically active users in the \textit{base} dataset and could be the result of employing COVID19 for political messaging.  However, this requires further investigation. Nonetheless the disparity implies that users in the \textit{politicised} dataset seem to be actively spreading their message in light of their frames.

\begin{figure}
    \centering
    \includegraphics[width=0.7\linewidth]{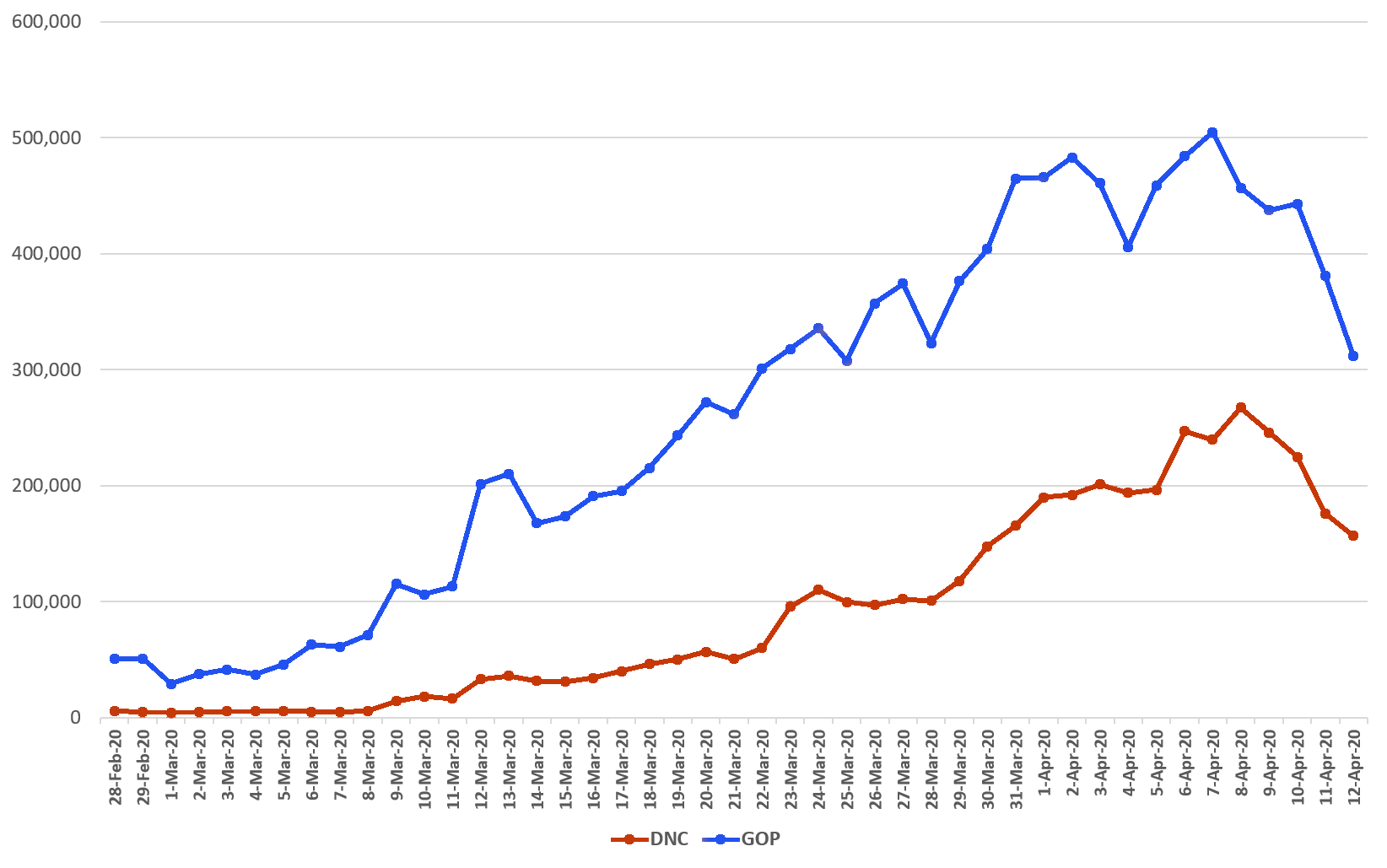}
    \caption{The number of daily tweets of DNC and GOP learning users.}
    \label{fig:dailyTweetsOfDNCGOPSupporters}
\end{figure}

\emph{\textbf{User Stance Detection}} Given the \textit{politicized} dataset filtered on COVID19 related terms, we wanted to determine which users politically lean towards the DNC or the GOP.  To do so, we employed an unsupervised stance detection method, which attempts to discriminate between users based on the accounts that they retweet \cite{darwish2019unsupervisedStance}. We elected to use this method because it was shown to produce nearly perfect user clusters. The method represents each user using a vector of the accounts that they retweeted. Then it computes the cosine similarity between users and projects them onto a two dimensional space using UMAP, which places the users in a manner where similar users are closer together and less similar users are further apart.  Next, projected users are clustered using mean shift clustering.  Of the 30k users, stance detection was able to assign 25,753 users to two main clusters and was unable to cluster the remaining users.  By inspecting the two clusters, the first cluster of size 17,689 users was clearly composed of DNC leaning users and the second cluster of 8,064 users, who were clearly leaning towards the GOP. Since less than 1\% of the tweets were authored before Feb. 28, we restrict our analysis to the period from Feb. 28 to April 12, 2020. Figure \ref{fig:dailyTweetsOfDNCGOPSupporters} plots the number of daily tweets for DNC and GOP supporters.  

\section{Frames}
\textbf{\textit{Identifying Frames.}} We wanted to determine the frames for both the \textit{sampled} and \textit{politicized} dataset.  For the \textit{politicised} dataset, we analyzed the top 100 most frequently used hashtags for pro-DNC and pro-GOP users.   Tables~\ref{tab:topHashtagGOP} and \ref{tab:topHashtagDNC} categorize the 100 most frequent hashtags (excluding COVID19 hashtags) by pro-GOP and pro-DNC users respectively.  In looking at these hashtags, we can better understand the frames of both groups.  For pro-GOP users (Table~\ref{tab:topHashtagGOP}), we can see that the key factors in framing are:  COVID19 is a conspiracy, blaming China, Trump reelection, liberal media credibility, Trump's effectiveness, party loyalty, social solidarity, and patriotism. For the pro-DNC users (Table~\ref{tab:topHashtagDNC}), the factors are: defeating Trump (anti-Trump/pro-DNC), Trump's ineptness, and social solidarity.  As is apparent from these factors, there are 3 overarching frames for both groups, namely: 
1. Assignment of blame: pro-GOP users blame conspiracies, China, and left-leaning media, while pro-DNC users blame Trump and the GOP; 
2. Support for party candidate(s):  pro-GOP users support Trump and GOP candidates, while pro-DNC users support DNC candidates; and 
3. Social messaging concerning the COVID19 lockdown.

Pro-GOP users had an additional frame that ``everything is OK'' (e.g. possible cure, holiday greetings), implicitly supporting that status quo.  Assignment of blame and support for party candidate(s) are associated with the Nov. 2020 election, which is both mentally accessible and relevant. They account for 77\% and 87\% of hashtag volume for pro-GOP and pro-DNC users respectively. 

For the \textit{sampled} dataset, we analyzed the top 200 most frequently hashtags, excluding COVID19 related hashtags.  As shown in Table~\ref{tab:sampledDatasetTopHashtags}, the main themes were: entertainment (e.g. games, music, sports), following the news, politics (e.g. supporting/opposing Trump/GOP/DNC), science and technology (e.g. big data, artificial intelligence, machine learning), motivational, conspiracy theories (e.g. QAnon), and other issues.  Unlike the \textit{politicized} dataset, the volume of election related themes was significantly lower (27\% of the total).  This implies that politically active users are not necessarily representative of the entire US Twitter user population.  However, their influence is visible in the general population.
\begin{table}[pt!]
    \centering
        \caption{Categorization of the 100 most frequent hashtags (excluding COVID19 hashtags) by pro-GOP users.}
        \tiny
    \begin{tabular}{p{2.6cm}|r|r|p{7cm}}
Category	&	Count	&	\%	&	Examples	\\ \hline
Conspiracy	&	231,132	&	17.8	&	\#FilmYourHospital, \#BillGates, \#Qanon (deep state conspiracy)	\\
Blame China	&	216,549	&	16.7	&	\#ChineseVirus, \#WuhanCoronaVirus \\
Trump support	&	135,602	&	10.4	&	\#MAGA (Make America Great Again), \#Trump2020, \#KAG2020 (Keep America Great)	\\
Anti liberal media	&	131,705	&	10.1	&	\#FakeNews, \#EnemyOfThePeople	\\
Anti DNC	&	117,760	&	9.1	&	\#DemocratsHateAmerica	\\
Pro GOP	&	88,395	&	6.8	&	\#walkaway, \#TCOT (Top Conservatives on Twitter) \\
News	&	60,620	&	4.7	&	\#BREAKING	\\
Praising Trump's actions	&	60,378	&	4.7	&	\#PaycheckProtectionProgram, \#AmericaWorksTogether	\\
COVID19 cure	&	56,420	&	4.3	&	\#Hydroxychloroquine	\\
Specific issues	&	51,629	&	4.0	&	\#Iran, \#FISA	\\
Social solidarity	&	38,820	&	3.0	&	\#InItTogether, 
\#SocialDistancing	\\
Holiday related	&	33,472	&	2.6	&	\#Easter, \#GoodFriday	\\
Conservative media	&	29,144	&	2.2	&	\#FoxNews	\\
voting conspiracy	&	19,401	&	1.5	&	\#VoterId, \#VoterFraud	\\
Patriortic sentiment	&	19,374	&	1.5	&	\#America	\\

    \end{tabular}
    \label{tab:topHashtagGOP}
\end{table}
\begin{table}[hbt!]
    \centering
        \caption{Categorization of the 100 most frequent hashtags (excluding COVID19 hashtags) by pro-DNC users.}
        \tiny
    \begin{tabular}{p{3cm}|r|r|p{6.8cm}}
Category	&	Count	&	\%	&	Examples	\\ \hline
anti Trump	&	1,105,341	&	56.6	&	\#TrumpVirus, \#FireTrump	\\
pro DNC	&	214,975	&	11.0	&	Biden2020, VoteBlue2020 	\\
blame Trump Administration	&	162,253	&	8.3	&	\#PPE (Personal Protective Equipment), WhereAreTheTests	\\
anti GOP	&	144,175	&	7.4	&	\#MoscowMitch, \#GOPGenocide	\\
social solidarity	&	126,468	&	6.5	&	\#StayHome, \#FamiliesFirst	\\
News	&	77,358	&	4.0	&	\#BREAKING	\\
voting issues	&	43,550	&	2.2	&	\#VoteByMail, \#WisconsinPrimary 	\\
Conservative media	&	32,619	&	1.7	&	\#BloodOnHisHandsHannity	\\
Liberal media	&	32,233	&	1.7	&	\#Madow, \#MSNBC	\\
Holocaust	&	14,285	&	0.7	&	\#Auschwitz	\\

    \end{tabular}
    \label{tab:topHashtagDNC}
\end{table}
\begin{table}[hbt!]
    \centering
    \caption{Categorization of the 200 most frequent hashtags (excluding COVID19 hashtags) by \textit{sampled} group.}
    \tiny
    \begin{tabular}{p{3cm}|r|r|p{6.8cm}}
Category	&	Count	&	\%	&	Examples	\\ \hline
Entertainment	&	128,803	&	33.3	&	\#AnimalCrossing, \#BTS, \#IHeartAwards, \#NFLDraft, \#Oscars	\\
News	&	30,952	&	8.0	&	\#Breaking, \#News	\\
Support Trump/GOP	&	23,386	&	6.1	&	\#MAGA, \#Trump2020, \#KAG	\\
Social solidarity	&	21,464	&	5.6	&	\#StayHome, \#SocialDistancing	\\
Science/Technology	&	21,138	&	5.5	&	\#AI, \#BigData, \#ML	\\
Business	&	19,837	&	5.1	&	\#Business, \#Marketing	\\
Democrats	&	16,656	&	4.3	&	\#DemDebate, \#Biden	\\
Attack Trump/GOP	&	13,467	&	3.5	&	\#TrumpVirus, \#Resist	\\
Support DNC	&	13,203	&	3.4	&	\#YangGang, \#Bernie2020	\\
Inspiration	&	12,770	&	3.3	&	\#Success, \#Wisdom, \#Leadership	\\
Conspiracy theories	&	11,964	&	3.1	&	\#Qanon, \#wwg1wga	\\
Republican	&	11,896	&	3.1	&	\#TCOT, \#GOP	\\
Issues	&	10,258	&	2.7	&	\#EarthDay, \#BlackHistoryMonth	\\
Generic	&	9,233	&	2.4	&	\#Love, \#Win, \#Truth	\\
Political	&	8,501	&	2.2	&	\#Impeachment, \#2020Census	\\
Health care	&	7,090	&	1.8	&	\#Health, \#MedicareForAll, \#MentalHealth	\\
China	&	6,734	&	1.7	&	\#China, \#Wuhan, \#CCP	\\
Foreign issues	&	6,409	&	1.7	&	\#WWIII, \#Iran	\\
Blame media	&	3,410	&	0.9	&	\#FakeNews	\\
Attack DNC	&	3,192	&	0.8	&	\#MeToo (accusing Biden of harassment)	\\
    \end{tabular}
    \label{tab:sampledDatasetTopHashtags}
\end{table}

\textbf{\textit{Analyzing Frames.}}  To further analyze the content of the tweets, we utilized multiple methods.  The first method involved identifying the most distinguishing features for both camps.  To do so, we employed a variation of the so-called valence score \cite{conover2011political}, which attempts to determine the distinctiveness of a given token, such as a hashtag or retweeted account, for a particular group. Valence for term $t$ in tweets of group $G_g$ is computed as follows:
\begin{equation}
    V(t,G_g) = 2 \frac{
        \frac{N(t,G_g)}{N(G_g)}}
        {\frac{N(t,G_g)}{N(G_g)} +\frac{N(t, G_{\neg{g}})}{N(G_{\neg{g}})}
    } - 1
\end{equation}

Where $N(t,G_g)$ is the frequency of term $t$ in the tweets of group $G_g$ (e.g. GOP tweets) and $N(t,G_{\neg{g}})$ is the frequency of the term in the other group (e.g. DNC), and $N(G_g)$ is the total number of all terms in group $G_g$. Valence values are bounded between -1 and 1, where -1 and 1 imply extreme disassociation or association respectively.  We split the range into five equal intervals, and 
we computed the valence scores for all hashtags and retweeted accounts for the \textit{politicized} dataset.  Tables~\ref{tab:topHashtags},~\ref{tab:topRetweetedAccounts}, and~\ref{tab:topURLs} list the most polarizing hashtags, retweets, and URLs, where the absolute value of the valence score is $\ge 0.6$, and they are ranked by the product the valence score and the log of the term frequency~\cite{stefanov2020predicting}.  The most distinguishing hashtags seem to reinforce the aforementioned frames.  On the GOP side, 6 out of 20 hashtags blame China (e.g. \#ChinaVirus), 4 promote conspiracy theories (e.g. \#QAnon), and 5 support Trump and his actions (e.g. \#KAG2020, \#USNSMercy). On the DNC side, 14 of the 20 hashtags in Table~\ref{tab:topHashtags} are targeted directly at Trump (e.g. \#TrumpVirus, \#Resist, \#25thAmendmentNow), 2 more hashtags implicitly attack Trump (\#PPE -- personal protective equipment, \#WhereAreTheTests), and yet 2 more attack the GOP (\#GOP, \#MoscowMitch).  The list of most prominent retweeted accounts is dominated by media, media personalities, politicians, and activists with conservative and liberal leanings respectively. Though all the tweets in the \textit{politicized} dataset mention COVID19, the most distinguishing hashtags and retweeted accounts reinforce political framing. As for the most distinctive URLs, pro-GOP users cited multiple government websites (e.g. White House), articles attacking democrats, China, and World Health Organization, and articles promoting the efficacy of hydroxychloroquine.  Pro-DNC users conversely attacked Trump and the GOP. 

\begin{table}[hbt!]
    \centering
        \caption{Top hashtags from GOP and DNC supporters from
        \textit{politicized} dataset}
        \tiny
    \begin{tabular}{l|r||l|r}
    \multicolumn{2}{c}{GOP Supporters} & \multicolumn{2}{c}{DNC Supporters} \\
Hashtag	&	Freq.	&	Hashtag	&	Freq.	\\\hline
China	&	29,993	&	TrumpVirus	&	37,229	\\
WuhanCoronaVirus	&	20,151	&	TrumpLiesAboutCoronavirus	&	26,776	\\
OANN	&	9,703	&	TrumpGenocide	&	20,502	\\
QAnon	&	9,309	&	SmartNews	&	18,747	\\
FakeNews	&	9,597	&	TrumpLiesAmericansDie	&	18,336	\\
CCPVirus	&	6,643	&	TrumpLiesPeopleDie	&	17,185	\\
ChineseVirus	&	7,914	&	TrumpOwnsEveryDeath	&	16,980	\\
Democrats	&	7,748	&	TrumpPandemic	&	15,741	\\
ChinaVirus	&	5,844	&	GOP	&	14,242	\\
Trump2020	&	6,420	&	TrumpVirusCoverup	&	14,782	\\
Hydroxycloroquine	&	5,582	&	StopAiringTrump	&	14,603	\\
WuhanVirus	&	5,427	&	TrumpLiedPeopleDied	&	14,260	\\
KAG2020	&	5,913	&	TrumpIsTheWORSTPresidentEVER	&	13,202	\\
USNSMercy	&	4,853	&	Resist	&	12,162	\\
WWG1WGA	&	4,591	&	FamiliesFirst	&	12,481	\\
USNSComfort	&	4,596	&	TrumpPlague	&	11,622	\\
WHO	&	5,912	&	MoscowMitch	&	10,133	\\
NEW	&	4,600	&	PPE	&	8,245	\\
KAG	&	6,133	&	WhereAreTheTests	&	8,099	\\
Iran	&	5,296	&	25thAmendmentNow	&	7,749	\\
    \end{tabular}
    \label{tab:topHashtags}
\end{table}

\begin{table}[hbt!]
    \centering
    \tiny
    \caption{Top retweeted accounts by pro-GOP and pro-DNC users in \textit{politicized} dataset}
    \begin{tabular}{l|r|p{3cm}||l|r|p{3cm}}
    \multicolumn{3}{c}{GOP Supporters} & \multicolumn{3}{c}{DNC Supporters} \\
Account	&	Count	&	Description	&	Account	&	Count	&	Description	\\ \hline
mitchellvii	&	100,680	&	pro-Trump host	&	funder	&	201,485	&	Scott Dworkin (liberal host)\\
TomFitton	&	79,787	&	president of conservative activist group	&	DrDenaGrayson	&	127,047	&	medical doctor	\\
jsolomonReports	&	73,168	&	conservative pundit	&	tedlieu	&	126,776	&	democratic congressman	\\
RealCandaceO	&	64,828	&	conservative author	&	kylegriffin1	&	115,110	&	MSNBC producer	\\
WhiteHouse	&	88,982	&	Official White House account	&	Yamiche	&	113,313	&	MSNBC contributor	\\
RealJamesWoods	&	57,697	&	actor/Trump supporter	&	JoeBiden	&	156,903	&	Biden's official account	\\
TrumpWarRoom	&	57,258	&	pro-Trump account	&	MSNBC	&	116,555	&	liberal media	\\
IngrahamAngle	&	57,642	&	FoxNews host	&	TeaPainUSA	&	79,974	&	liberal account	\\
marklevinshow	&	40,487	&	conservative radio host	&	joncoopertweets	&	75,994	&	democratic politician	\\
gatewaypundit	&	37,482	&	conservative newsletter	&	SethAbramson	&	72,541	&	Newsweek columnist	\\
BreitbartNews	&	37,504	&	conservative media	&	JoyAnnReid	&	73,680	&	MSNBC corespondent	\\
DonaldJTrumpJr	&	46,172	&	Trump's son	&	maddow	&	79,309	&	MSNBC host	\\
SaraCarterDC	&	33,207	&	FoxNews host	&	atrupar	&	68,202	&	Vox journalist	\\
catturd2	&	30,913	&	pro-Trump account	&	realTuckFrumper	&	57,377	&	liberal newsletter	\\
charliekirk11	&	32,510	&	president of conservative activist group	&	mmpadellan	&	54,389	&	liberal blogger	\\
seanhannity	&	36,724	&	FoxNews host	&	CREWcrew	&	53,963	&	liberal thinktank	\\
DailyCaller	&	25,781	&	FoxNews host	&	MollyJongFast	&	52,534	&	Daily Beast editor	\\
FLOTUS	&	30,819	&	First Lady's official account	&	washingtonpost	&	67,127	&	liberal media	\\
TeamTrump	&	26,864	&	Official Trump campaign account	&	Amy\_Siskind	&	49,278	&	liberal activist	\\
JackPosobiec	&	24,586	&	alt-right activist/conspiracy theorist	&	eugenegu	&	49,639	&	medical doctor	\\

    \end{tabular}
    \label{tab:topRetweetedAccounts}
\end{table}

\begin{table}[hbt!]
    \centering
    \caption{Top cited URLs from pro-GOP and pro-DNC users in \textit{politicized} dataset.}
    \tiny
    \begin{tabular}{r|p{11cm}}
Count	&	URL	\\ \hline
\multicolumn{2}{c}{GOP Supporters} \\ \hline
6,570	&	\url{http://CoronaVirus.gov}	\\
2,803	&	\url{https://nypost.com/2020/04/09/senate-dems-block-250-billion-for-coronavirus-small-business-loans/}	\\
2,606	&	\url{http://WWW.GEORGE.NEWS}	(conservative news site)\\
2,511	&	\url{https://www.wsj.com/articles/world-health-coronavirus-disinformation-11586122093}	\\
2,482	&	\url{https://www.whitehouse.gov/briefings-statements/coronavirus-guidelines-america/}	\\
2,456	&	\url{http://45.wh.gov/RtVRmD}	(White House Press Conference)\\
2,055	&	\url{https://www.freep.com/story/news/local/michigan/detroit/2020/04/06/democrat-karen-whitsett-coronavirus-hydroxychloroquine-trump/2955430001/}	\\
2,048	&	\url{https://www.foxnews.com/politics/after-mocking-trump-promoting-hydroxychloroquine-media-acknowleges-might-treat-coronavirus}	\\
2,027	&	\url{https://www.usatoday.com/story/opinion/2020/04/07/time-put-china-lockdown-dishonesty-amid-coronavirus-pandemic-crisis-column/2954433001/}	\\
2,000	&	\url{https://thegreggjarrett.com/coronavirus-crisis-americans-fear-the-lockdown-more-than-the-virus/}	\\
\hline
\multicolumn{2}{c}{DNC Supporters} \\ \hline
5,252	&	\url{https://www.latimes.com/politics/story/2020-04-07/hospitals-washington-seize-coronavirus-supplies}	\\
4,370	&	\url{https://secure.actblue.com/donate/coronavirus-liar-video}	\\
4,062	&	\url{https://www.nytimes.com/2020/04/11/us/politics/coronavirus-red-dawn-emails-trump.html}	\\
3,600	&	\url{https://www.queerty.com/fox-news-officially-sued-peddling-coronavirus-misinformation-20200406}	\\
3,485	&	\url{https://www.nytimes.com/interactive/2020/03/25/opinion/coronavirus-trump-reopen-america.html}	\\
3,392	&	\url{https://www.npr.org/sections/coronavirus-live-updates/2020/04/08/829955099/federal-support-for-coronavirus-testing-sites-end-as-peak-nears}	\\
3,340	&	\url{https://www.nytimes.com/2020/04/02/opinion/jared-kushner-coronavirus.html}	\\
3,417	&	\url{https://www.thedailybeast.com/sen-kelly-loeffler-dumped-millions-in-stock-after-coronavirus-briefing}	\\
2,978	&	\url{https://www.politico.com/news/2020/03/31/trump-obamacare-coronavirus-157788}	\\
3,201	&	\url{https://www.state.gov/the-united-states-announces-assistance-to-combat-the-novel-coronavirus/}	\\
    \end{tabular}
    \label{tab:topURLs}
\end{table}

The second method involved using DocuScope \cite{Ishizaki2011}, which is a text visualization and analysis environment specifically designed to carry out rhetorical research with language and text. The DocuScope dictionary was developed based on David and Brian Butler's theoretical work on rhetoric \cite{Kaufer1996} and their applied work in representational theories of language \cite{Kaufer2000}.  DocuScope is based on more than 60 million English linguistic patterns that map textual segments to fine-grained rhetorical effects which allows analysts to engage in deep cultural interpretation and extract sociocultural trends from text. We analyzed the tweets from both camps using DocuScope, and we used the output in two ways.  First, we looked at the tweets that mention specific keywords, namely: China, Trump, Republican, Democrat, media, and conspiracy, which are strongly associated with the aforementioned frames.  We identified the rhetorical devices and ranked them using the the product of the valence score and the log of their frequency. Table~\ref{tab:rhetroicalDevicesTargets} lists the top 5 most associated rhetorical devices that are used in conjunction with each target.  Again, we can see that the rhetorical devices are used in a manner that is consistent with the frames.  GOP supporters are blaming China (source of virus, no human rights), blaming Democrats (for ``phony'' impeachment, supporting illegal immigrants), attacking media (as tool of globalists and leftist, being a sham), touting a potential drug (hydroxychloroquine), and promoting conspiracy theories (e.g. COVID19 targets elderly GOP supporters). As for DNC supporters, they are used to blame Trump (as incompetent, failure), blame the GOP (for voter suppression, helping the rich), call for help to defeat republicans in elections, and accusing Trump and conservative media of promoting conspiracy theories.

\begin{table}[bht!]
    \centering
        \caption{Rhetorical devices used in conjunction with: China, Trump, Republicans, Democrats, Media, and Conspiracies}
    \tiny
    \begin{tabular}{p{2.1cm}|p{3.7cm}||p{2.1cm}|p{3.7cm}}
        \multicolumn{2}{c||}{GOP Supporters} & \multicolumn{2}{|c}{DNC Supporters} \\
    \multicolumn{4}{c}{China} \\
Negative/ Bogus, Phony & distracted from China by \textbf{phony} impeachment	&	Narrative/ Forgiveness 	&	Trump should ask \textbf{forgiveness} for incompetence	\\ \hline
Inquiry/ Curiosity, Intrigue	&	\textbf{intriguing}	if China had vaccine &	Investigate/ Detect	&	US gov failed despite \textbf{detecting}	virus early\\ \hline
Description/ Objects Gate	&	\textbf{Gates} wants to vaccine and track people	&	Negative/ Incompetent	&	complaint that Trump is an \textbf{incompetent idiot}	\\ \hline
Inquiry/ Investigate	&	Complaint Congress has \textbf{been investigating} Trump	&	Strategic/ Goals, Aim	&	Trump ended funds \textbf{aimed at} early warning	\\ \hline
Reasoning/ Ethics, Human	&	China lacks \textbf{human rights}	&	Negative/ Trickery	&	Trump calling virus a \textbf{hoax}	\\ \hline
\hline
    \multicolumn{4}{c}{Trump} \\
    Information/ Left	&	\textbf{the left} mad over Jared Kushner	&	Negative/ Incompetence	&	Trump \textbf{incompetence}	\\ \hline
Political Ideology/ Democrat	&	\textbf{Democratic state} Rep says that Hydroxychloroquine saved her life	&	Negative/ Firing	&	Trump \textbf{fired} the pandemic response team	\\ \hline
Negative/ Restrict	&	Trump \textbf{restricting} travel to China	&	Negative/ Lapse, Failure	&	\textbf{failure of} Trump over COVID19	\\ \hline
Negative/ Trickery, Sham	&	calling the impeachment a \textbf{sham}	&	Negative/ Bungle	&	Trump \textbf{bungled} testing	\\ \hline
Political Ideology/ Left wing	&	attacking \textbf{left-wing} media smear	&	Narrative/ Turn Break	&	\textbf{major breaking}: Trumped blocked testing	\\ \hline
\hline
    \multicolumn{4}{c}{Republicans} \\
Description/ Drug	&	\textbf{hydroxychloroquine} is effective	&	Positive/ Join our	&	\textbf{join our} fight against Republicans	\\ \hline
Strategic/ Goals, Target	&	COVID19 \textbf{targets} elderly (mostly republicans)	&	Interactive/ Request help	&	\textbf{help} defeat Republicans	\\ \hline
Negative/ Burdens	& \multirow{3}{3.7cm}{Republicans calling models \textbf{used for} \textbf{projection} conflicted and cause undue \textbf{burden} on Americans}	&	Negative/ Suppress	&	GOP accused of voter \textbf{suppression}	\\ 
\cline{1-1} \cline{3-4}
Future/ Projection	&		&	Information/ Expunge, Remove	&	Republicans \textbf{could have removed} Trump but did not	\\ 
\cline{1-1} \cline{3-4}
Narrative/ Used to	&		&	Information/ Report cost	&	Accusing GOP \textbf{used} \$500B to help rich donors	\\ \hline
\hline
    \multicolumn{4}{c}{Democrats} \\
Description/ Drug	&	\multirow{2}{3.7cm}{Democratic representative praises Trump \textbf{touted} drug (\textbf{hydroxychloroquine})}	&	Positive/ Join our	&	\multirow{3}{3.7cm}{\textbf{help} us/\textbf{join our} effort to defeat \textbf{Republicans}}	\\ \cline{1-1} \cline{3-3}
Strategic/ Persuasion, Touting	&		&	Interactive/ Request help	&		\\ \cline{1-3} 
Negative/ llegal	&	Accusing DNC of demanding relief for \textbf{illegal immigrants}	&	Public/ Political Ideology/ Republican	&		\\ \hline
Description/ Cards	&	DNC fighting for COVID19 \textbf{card} but against voted ID	&	Public/ Legislature	&	\multirow{2}{3.7cm}{Accusing WI \textbf{legislature} and \textbf{court} of conspiring to help GOP in election}	\\ \cline{1-3}
Positive/ Tremendous	&	relief bill \textbf{tremendous opportunity} that DNC missed	&	Public/ Law/ Courts	&		\\ \hline
\hline
    \multicolumn{4}{c}{Media} \\
Information/ Left	&	the \textbf{left} leaning media refuses to air Trump's press conference	&	Narrative/ Turn Break	&	\textbf{major breaking news:} Trump spreading misinformation	\\ \hline
Character/ Globalist	&	media is manipulated by \textbf{globalists}	&	Negative/ Incompetence	&	media calls Trump's actions \textbf{incompetence}	\\ \hline
Positive/ Loyalty	&	Trump exposes media -- follow \textbf{patriots} instead	&	Negative/ Squander	&	Trump \textbf{squandered} his credibility	\\ \hline
Negative/ Trickery, Sham	&	media coverage is a \textbf{sham}	&	Negative/ Bungle	&	Media exposes \textbf{bungled} COVID19 response	\\ \hline
Negative/ Bogus, Phony	&	media concentrated on \textbf{phony} impeachment	&	Negative/ Act Aggressive, Fire	&	Trump \& GOP spreading misinformation, while Trump \textbf{fired} response team	\\ \hline
\hline
    \multicolumn{4}{c}{Conspiracy} \\
Negative/ Fail to work	&	\multirow{5}{3.7cm}{Fauci thinks \textbf{Hydroxychloroquine} \textbf{doesn't work}. \textbf{Beginning to think} he is either an \textbf{idiot} or \textbf{lying to} us.}	&	Character/ Personality	&	\textbf{Murdoch} (FoxNews) spreading conspiracy theories	\\ \cline{1-1} \cline{3-4}
FirstPerson/ Anger	&		&	Updates/ BreakingNews	&	\textbf{breaking news:} lawsuit against Trump, Murdoch, and FoxNews coming \textbf{soon}	\\ \cline{1-1} \cline{3-3}
Negative/ Stupid person	&		&	Narrative/ Time shift, soon	&		\\ \cline{1-1} \cline{3-4}
Description/ Drug	&		&	Description/ Mask	&	Trump accuses nurses of stealing \textbf{masks}	\\ \cline{1-1} \cline{3-4}
Narrative/ Embark	&		&	Public/ Administer	&	Trump \textbf{administration} spreading conspiracy theories	\\ 

    \end{tabular}
    \label{tab:rhetroicalDevicesTargets}
\end{table}

\begin{table}[pt]
    \centering
        \caption{Top positive and negative rhetorical devices for GOP and DNC supporters}
    \tiny
    \begin{tabular}{p{2.1cm}|p{3.7cm}||p{2.1cm}|p{3.7cm}}
    \multicolumn{2}{c||}{GOP Supporters} & \multicolumn{2}{|c}{DNC Supporters} \\
    \multicolumn{4}{c}{Positive} \\
Positive/ High quality	&	food supply workers delivering \textbf{high quality} products	&	Positive/ Relieve, Acquittal	&	Trump hid COVID19 to secure impeachment \textbf{acquittal}	\\ \hline
Positive/ Guardian, Prophylactic	&	India approved Hydroxycloroquine as \textbf{prophylactic}	&	Positive/ Greater good	&	Trump not working for the \textbf{greater good}	\\ \hline
Positive/ Outstanding	&	Fauci defies Trump, calls WHO  boss '\textbf{outstanding} person'	&	Positive/ Empathy, Hear one out	&	\textbf{hear me out:} Trump failed	\\ \hline
Positive/ Tolerant	&	\textbf{``tolerant''} left cause violence	&	Positive/ Speak truth to power	&	Trump pursuing public servants who \textbf{speak truth to power}	\\ \hline
Information/ Relieve, Vindicate	&	COVID19 \textbf{vindicates} Trump's immigration policy	&	Information/ Relieve,Exculpatory	&	Trump's actions are not \textbf{exculpatory} (but damming) 	\\ \hline
    \multicolumn{4}{c}{Negative} \\
Negative/ Illegal	&	DNC wants to help \textbf{illegal aliens}	&	Negative/ Act aggressive, Fire & Trump \textbf{fired} pandemic team	\\ \hline
Negative/ Resist, Deny	&	\textbf{unrelated} deaths blamed on COVID19	&	Negative/ Incompetence	&	Trump is \textbf{incompetent}	\\ \hline
Negative/ Bogus, Phony	&	complaint about \textbf{phony} impeachment	&	Negative/ Unforgivable	&	Trump's inaction \textbf{unforgivable}	\\ \hline
Negative/ Trickery, Sham	&	media is a sham	&	Negative/ Mother fucker	&	Trump is \textbf{…}	\\ \hline
Negative/ Fudge data	&	CDC \textbf{fudging the numbers} (for COVID19)	&	Negative/ Fucked up	&	Trump \textbf{…} COVID19 response	\\ \hline

    \end{tabular}
    \label{tab:PositiveNegativeDocuScope}
\end{table}

The second way we used DocuScope output is that we inspected the top 5 positive and negative devices for both groups and ranked them using the product of the valence and log of the frequency.  As shown in Table~\ref{tab:PositiveNegativeDocuScope}, the devices are consistent with the aforementioned frames.  While positive devices were often used by GOP supporters to express support for Trump, the positive devices were used by DNC supporters to formulate attacks against Trump (e.g. going after public servants, seeking acquittal from crimes).  Usage of negative devices were used in attacking media, CDC, and Democrats by GOP supporters, and for attacking Trump by the DNC supporters.  \\

\textbf{\textit{Discussion of Frames.}} Framing is autological. The purpose of frames is to ``serve as bridges between elite discourse about a problem or issue and popular comprehension of that issue'' \cite{nelson1997toward}. Political groups dedicate a substantial amount of time towards regulating not just the information that is shared but also how it is presented~\cite{nelson1997toward}. Translated to Twitter, framing has become a vital tool for both the GOP and the DNC, in order for them to relay (or spin) information to the members of their parties and any users who display interest. Using Twitter for a political purpose, Democrats have reinforced a negative frame around Donald Trump, weaving the COVID19 pandemic into the picture for users that are avidly reading and absorbing the hashtags and the tweets. As for Republicans, there is a more positive form of framing that from the start has been built to evoke fellow party members to protect and support Trump. As a result, the COVID19 blame has been continuously shifted towards conspiracies theories (e.g. deep state, liberal media, etc.) and China as the source of the virus. Using both a blame frame and a support frame, COVID19 is framed as a political issue.  These frames are somewhat predictable, because they reinforce the sentiment of each party, influencing users to continue to support what they already believe. According to framing theory, it is believed that ``people draw their opinions from the set of available beliefs stored in memory. Only some beliefs become accessible at a given moment. Out of the set of accessible beliefs, only some are strong enough to be judged relevant or applicable to the subject at hand'' \cite{chong2007framing}. In this case, the framing that is reinforced is relevant to the upcoming presidential campaign and has an emotional inclination due to factors such as COVID19.  Both Democrats and Republicans have a prominent following on Twitter, and political framing can be observed in the top tweets by both party supporters. The purpose of these tweets and hashtags, as well as URLs and links, are to present to both audiences why Trump is fit or unfit to be reelected. 

\begin{table}[pt]
    \centering
        \tiny
        \caption{Top tweets from GOP and DNC supporters from \textit{politicized} dataset}
    \begin{tabular}{c|p{11.3cm}}
        Freq. & Text \\ \hline
        \multicolumn{2}{c}{GOP Supporters} \\ \hline
         5,008 & RT @realDonaldTrump: .@OANN  A key CoronaVirus Model is now predicting far fewer deaths than the number shown in earlier models. That’s because the American people are doing a great job. Social Distancing etc. Keep going! \\ \hline
        4,998 & RT @realDonaldTrump: So now the Fake News @nytimes is tracing the CoronaVirus origins back to Europe, NOT China. This is a first! I wonder what the Failing New York Times got for this one? Are there any NAMED sources? They were recently thrown out of China like dogs, and obviously want back in. Sad! \\ \hline
        3,795 & RT @DonaldJTrumpJr: Take a look back and RT what the liberal media and Joe Biden said about President Trump's aggressive, early response to the \#coronavirus.   Thank God @realDonaldTrump is the one in charge during this scourge! https://t.co/d6OtmWrGmb \\ \hline
        \multicolumn{2}{c}{DNC Supporters} \\ \hline
        6,435	&   RT @tribelaw: What if we were to learn that Trump suppressed scary information re COVID19 (and the needed federal response) in January to postpone the economic turndown until it could no longer endanger his Senate acquittal? Retweet if you wouldn’t be surprised by his making that tradeoff.\\ \hline
        6,207	&   RT @RepMaxineWaters: Trump, you incompetent idiot! You sent 18 tons of PPE to China early but ignored warnings \& called COVID19 concerns a hoax. You've endangered doctors, nurses, aids, orderlies, \& janitors - all risking their lives to save ours. Pray 4 forgiveness for the harm that you're\\ \hline
        6,114	&   RT @BrianKarem: Lies. On 2/28/20 on the South Lawn I ASKED you about the W.H.O. telling us the risk for COVID19 had increased. You blew off the question to tell us about Your rally that night. You blew it. Not the W.H.O. @realDonaldTrump https://t.co/DHhSgURGRt \\
    \end{tabular}

    \label{tab:topTweets}
\end{table}

From inspecting the top retweeted tweets (Table~\ref{tab:topTweets}), 
while the top DNC tweets are pretty self-explanatory, the emotions behind them are negatively geared against Trump. 
They are meant to demonstrate that Trump is not suitable to remain president, because he showed misfeasance or perhaps malfeasance in handling the pandemic.  
On the opposite end, the framing theory is seen in full effect for Republicans as a positive source of influence to show that Trump is a suitable president both now and in the future. For example, the tweet by @DonaldJTrumpJr (Table~\ref{tab:topTweets}) 
is directly related to COVID19 and frames the perspective that Trump is a fit president that handled the pandemic properly from the beginning. The other two tweets by @realDonaldTrump are expressing hope that everything will OK and attacking the media as fake news. \newline 
Aside from tweets that are direct, frames are regularly reinforced by both Democrats and Republicans through routine hashtags and URLS. The top hashtags from both parties in the period of study include Democratic hashtags such as \#TrumpPlague \#TrumpGenocide, and \#TrumpLiesAboutCoronavirus. As for Republican influenced hashtags, these include hashtags such as \#OANN and \#WuhanCoronaVirus. 
Correlating with the framing theory, there is focus on the pandemic and it is framed to fit a certain understanding that is different for each party. For Democrats, it is directly blaming Trump for COVID19. As for Republicans, it is shifting the blame from Trump to conspiracies and China, which is where the virus originated. Adding emphasis to particular parts of issues allows for politicians and political parties to create an mental association and to put them in specific reference frames \cite{johnson2017modeling}. There is also usage of the hashtag \#OANN, which refers to the conservative news channel (One America News Network). Since COVID19 updates are provided by the media, using the hashtag \#OANN directly connects users to Republican influenced news.  Through framing, OANN presents opinion and information from a Conservative perspective. \\
The top URLs from both Republican and Democratic Twitter accounts are primarily focused on two categories: they are COVID19 and the presidential election (Table~\ref{tab:topURLs}). For Republican URLs, the top URLs include the official government website on COVID19 (CoronaVirus.gov), which displays pandemic updates, ways to protect yourself, and guidelines to re-open America. This influences the notion that Trump is acting accordingly as the president. There are also multiple URLs that link to websites such as USA Today and The Wall Street Journal, which focus on blaming China for not sharing early on how threatening COVID19 is. 
The usage of these URLs frames the idea that the blame and the emotions that come with the reaction to blame are to be focused on China.\\
From a Democratic perspective, the use of URLs as a source of political framing is used to influence users to blame Trump for his inability to contain the spread of COVID19. 
The URLs include a link to the Democratic Coalition’s fundraiser to create ads aimed at exposing Trump’s lies about COVID19. There are also links to articles in media such as Foreign Policy, Washington Post, and New York Times, which echo the sentiment of Trump’s leadership failure both before COVID19 reached the United States and after it became a pandemic. Through political framing, the URLs are used to foreshadow President Trump’s leadership efforts.  Democrats directly set the agenda that Trump’s inadequacy to act early, in order to prevent COVID19, renders him unfit for the upcoming presidency. 
 
Through the usage of rhetorical devices, framing is reinforced to support positive or negative thoughts. The most popular Republican supported categories include, but are not limited to, conspiracy, blame China, praising Trump's actions, and Pro GOP. Hashtags that support President Trump evoke a positive rhetoric. For example, hashtags that directly relate to Trump are \#AmericaWorksTogether and \#KAG (Keep America Great). Examples of these hashtags strengthen the sentiment of America's togetherness and the continued need for it to remain great. Contrarily, the rhetoric used towards China displays a more negative connotation. This can be seen through hashtags such as \#ChineseVirus and \#WuhanCoronaVirus. These crafted phrases are used to frame the perception that China and COVID19 are synonymous with one another. 
 
Aside from blame and support frames that are prominently used by both parties, a social solidarity frame is visible for both parties as reflected by the hashtags \#SocialDistancing, \#StayHome and \#FamiliesFirst.  
As a rhetorical device, these hashtags illustrate that the GOP and DNC claim to prioritize the worth of people's health and families' financial well being. 

\section{Conclusion}
In this paper, we illustrate how framing can serve to shape the public discourse on an issue.  Specifically, we show how politically active supporters of the Democratic and Republican parties distinctly frame the COVID19 pandemic as a political issue -- as opposed to it solely being a public health issue. We use unsupervised stance detection to effectively identify the supporters of each party, and we analyze their most distinctive hashtags, retweeted accounts, URLs, and rhetorical devices.  In doing so, we are able to identify the frames that underlie their discourse.  Supporters of both parties crafted two primary frames, namely: a blame frame, where GOP supporters blame China and conspiracies, and DNC supporters blame Trump and the GOP; and a support frame, where each supports the candidate(s) of their respective parties. 
These frames dominate the discourse of both groups, thus revealing that support and blame prominently influence the political sentiments attached to COVID19.

\bibliographystyle{splncs04}
\bibliography{references}

\end{document}